\documentclass[prl,reprint,showpacs,superscriptaddress,bibnotes,twoside]{revtex4-1}

\usepackage{amsmath}
\usepackage{amssymb}
\usepackage{graphicx}

\begin{document}

\title{Non-Conventional Anderson Localization in Bilayered Structures with Metamaterials}

\author{E.~J.~Torres-Herrera}
\affiliation{Instituto de F{\'i}sica, Universidad Aut\'onoma de Puebla,
Apt. Postal J-48, Puebla, Pue., 72570, M\'exico}

\author{F.~M.~Izrailev}
\affiliation{Instituto de F{\'i}sica, Universidad Aut\'onoma de Puebla,
Apt. Postal J-48, Puebla, Pue., 72570, M\'exico}
\affiliation{NSCL and Dept. of Physics and Astronomy, Michigan State
University - East Lansing, Michigan 48824-1321, USA}

\author{N.~M.~Makarov}
\email{makarov.n@gmail.com}
\affiliation{Instituto de Ciencias, Universidad Aut\'{o}noma de Puebla, Priv. 17 Norte No 3417, Col. San Miguel Hueyotlipan, Puebla, Puebla, 72050, M\'exico}

\date{\today}

\begin{abstract}
We have developed an approach allowing us to resolve the problem of non-conventional Anderson localization emerging in bilayered periodic-on-average structures with alternating layers of right-handed and left-handed materials. Recently, it was numerically discovered that in such structures with weak fluctuations of refraction indices, the localization length $L_{loc}$ can be enormously large for small wave frequencies $\omega$. Within the fourth order of perturbation theory in disorder, $\sigma^2 \ll 1$, we derive the expression for $L_{loc}$ valid for any $\omega$. In the limit $\omega \rightarrow 0$ one gets a quite specific dependence, $L^{-1}_{loc} \propto \sigma ^4 \omega^8$. Our approach allows one to establish the conditions under which this effect can be observed.

\end{abstract}


\maketitle

\emph{Introduction.} -- The remarkable success in nano- and material science has led to a burst of interest to the study of wave propagation and electron transport in periodic one-dimensional (1D) systems (see, e.g., Refs.~\cite{MS08} and references therein). Among others, the systems of a particular interest are {\it bilayered structures} in optics and electromagnetics, or semiconductor superlattices, as well as alternating quantum wells and barriers in electronics \cite{all}. The interest to this subject is due to a possibility to create devices with prescribed properties of transmission and/or reflection.

As is shown in recent studies \cite{SSS01,Go06,Ao07,Ao10,Ao10a,Mo10}, the new perspectives are related to specific optic properties of metamaterials that can be embedded in periodic structures. A particular system that has attracted much attention, consists of two arrays of $a,b-$layers with equal widths, $d_a=d_b$, one of which is of right-handed (RH) material with positive refraction index, $n_a>0$, and the other is of left-handed (LH) material with negative index, $n_b<0$. In such periodic structure with equal $n_a=|n_b|$ the phase shift of wave function gained in $a-$layer is canceled by the subsequent shift in next $b-$layer. Therefore, the total phase shift vanishes after passing $N$ unit $(a,b)$ cells, that, together with perfect transmission, results in a kind of ``non-visible" structure.

However, even a small amount of disorder can destroy the above ideal picture due to an emergence of localization \cite{Go06}. To great surprise, the numerical data \cite{Ao07} obtained for the model with weakly disordered refraction indices $n_a$ and $n_b$, have demonstrated very fast divergence of localization length, $L_{loc} \sim\omega ^{-6}$, for $\omega \rightarrow 0$. In order to explain such a non-conventional Anderson localization, in Refs.\cite{Ao07,Ao10,Ao10a} it was suggested that the RH-LH structures can not be fully described via the Lyapunov exponent $\lambda$ as an inverse localization length. Recently, this study was extended to the case of off-axis incidence, for which the transport characteristics have been investigated in great detail \cite{Ao10a}.

In Ref.~\cite{IM09} another type of disorder was analyzed according to which both widths $d_a$ and $d_b$ are slightly fluctuating. Specifically, the general case of correlated disorder was analytically studied by assuming any kind of statistical correlations in widths of $a$ and $b-$layers, as well as the inter-correlations between the two widths. It was shown that for any ratio between $d_a$ and $d_b$ the localization length is governed by the unique expression, no matter whether the $a,b-$layers are both of normal materials (RH-RH array) or with alternating right-handed and left-handed materials (RH-LH array). The same statement stems from the analysis of RH-RH and RH-LH arrays with fluctuating refraction indices \cite{IMT10}, however, only when $d_a \neq d_b$. These results indicate that the abnormal localization emerges in a very specific situation, and not only due to inclusion of LH-arrays into the structure.

In order to shed light on this problem, in Ref.~\cite{TIM11} the RH-LH model with equal widths $d_a=d_b$ has been analyzed with the use of a powerful method developed in Ref.~\cite{IRT98,IMT10}. It was found that the phase of wave, propagating through the RH-LH array with fluctuating refraction indices, is described by a highly non-uniform distribution. This effect is somewhat similar to that arising in standard tight-binding models close to band edges (see, e.g. review \cite{IKM11} and references therein). Further analysis of this fact has led to a remarkable result \cite{TIM11,IKM11}: the Lyapunov exponent vanishes in the quadratic approximation in disorder parameter. Nevertheless, the problem has remained open since the evaluation of next order terms for the Lyapunov exponent was confronted with severe technical difficulties.

In this Letter we suggest the method allowing one to resolve the above problem of non-conventional Anderson localization. We derive the expression for the Lyapunov exponent within the fourth order perturbation theory, which is valid for {\it any} value of frequency $\omega$ and leads to the asymptotic expression $L_{loc} \propto \sigma^{-4} \omega^{-8}$ for small $\omega$. Our approach elucidates the mechanism that is responsible for such an abnormal localization, and allows one to specify the conditions under which it emerges.

\emph{Model.} -- The model describes the propagation of electromagnetic waves of frequency $\omega$ through an infinite array of two alternating $a$ and $b$ layers (slabs), respectively specified by the dielectric permittivity $\varepsilon_{a,b}$, magnetic permeability $\mu_{a,b}$, refractive index $n_{a,b}=\sqrt{\varepsilon_{a,b}\mu_{a,b}}$, impedance $Z_{a,b}=\mu_{a,b}/n_{a,b}$ and wave number $k_{a,b}=\omega n_{a,b}/c$. We consider two systems: the \emph{homogeneous} stack when both $a$ and $b$ layers are made of right-handed optic materials, and \emph{mixed} stack when $a$ layers contain RH-material while $b$ layers are of left-handed material. Following Ref.~\cite{Ao07}, we admit that the compositional disorder is incorporated via dielectric constants $\varepsilon_{a,b}$ only,
\begin{subequations}\label{PRL-nmu}
\begin{eqnarray}
&&n_a(n)=1+\eta_a(n),\quad\mu_a=1,\label{PRL-nmua}\\
&&n_b(n)=\pm[1+\eta_b(n)],\quad\mu_b=\pm1.\label{FNT-nZb}
\end{eqnarray}
\end{subequations}
Here $n$ enumerates the $n$th unit $(a,b)$ cell. The upper/lower sign stands for the RH/LH material respectively. Without disorder, $\eta_{a,b}(n)=0$, the RH-RH structure is just the air without any stratification, while the RH-LH array represents the \emph{ideal} mixed stack ($\varepsilon_a=\mu_a=1$, $\varepsilon_b=\mu_b=-1$) with perfectly matched slabs ($Z_a=Z_b=1$). All slabs have equal thicknesses, $d_a=d_b=d/2$, thus the size of any $(a,b)$ cell is $d$.

We specify that delta-correlated sequences $\eta_{a,b}(n)$ have zero average, $\langle\eta_{a,b}(n)\rangle=0$, and variance $\sigma^2$,
\begin{equation}\label{PRL-CorrDef}
\langle\eta_a(n)\eta_b(n')\rangle=\sigma^2\delta_{ab}\delta_{nn'}\,.
\end{equation}
Here $\langle ... \rangle$ stands for the average along arrays, which is assumed to be equivalent to ensemble average. Numerically, to generate $\eta_a(n), \eta_b(n')$ we use flat distribution, however our expressions are given in general form.

\emph{Basic relations.} -- As is shown in Refs.\cite{IM09,IMT10,TIM11}, ~the model can be written in the form of area-preserving map,
\begin{equation}\label{PRL-mapQP}
Q_{n+1}=A_nQ_n+B_nP_n,\quad P_{n+1}=-C_nQ_n+D_nP_n.
\end{equation}
Here the ``coordinate" $Q_n$ and ``momentum" $P_n$ refer to the wave function and its normalized derivative, respectively, taken at left-hand edge of the $n$th unit $(a,b)$ cell. The factors $A_n$, $B_n$, $C_n$, $D_n$ read
\begin{subequations}\label{PRL-ABCDn}
\begin{eqnarray}
A_n&=&\cos\varphi_a\cos\varphi_b-Z_a^{-1}Z_b\sin\varphi_a\sin\varphi_b,\\
B_n&=&Z_a\sin\varphi_a\cos\varphi_b+Z_b\cos\varphi_a\sin\varphi_b,\\
C_n&=&Z_a^{-1}\sin\varphi_a\cos\varphi_b+Z_b^{-1}\cos\varphi_a\sin\varphi_b,\\
D_n&=&\cos\varphi_a\cos\varphi_b-Z_aZ_b^{-1}\sin\varphi_a\sin\varphi_b.
\end{eqnarray}
\end{subequations}
For non-zero disorder the coefficients \eqref{PRL-ABCDn} depend on the cell number (discrete ``time") $n$, due to randomized refractive indices \eqref{PRL-nmu} entering {\it both} the impedances $Z_{a,b}(n)$ and phase shifts $\varphi_{a,b}(n)$,
\begin{subequations}\label{PRL-phi-ab}
\begin{align}
\varphi_{a}(n)&=k_a(n)d/2=\varphi[1+\eta_{a}(n)],\\
\varphi_{b}(n)&=k_b(n)d/2=\pm\varphi[1+\eta_{b}(n)],\quad\varphi=\omega d/2c.
\end{align}
\end{subequations}

It is useful to pass to polar coordinates $R_n$ and $\theta_n$ by the standard transformation,
\begin{equation}\label{PRL-QP-RTheta}
Q_n=R_n\cos\theta_n,\qquad P_n=R_n\sin\theta_n.
\end{equation}
The map in the radius-angle presentation gets the form
\begin{subequations}\label{PRL-mapRTheta}
\begin{eqnarray}
&&\left(R_{n+1}/R_n\right)^{-2}=d\theta_{n+1}/d\theta_n,\label{PRL-mapR-def}\\
&&\theta_{n+1}=\arctan\left[\frac{-C_n+D_n\tan\theta_n}{A_n+B_n\tan\theta_n}\right]\,.\label{PRL-mapTheta-def}
\end{eqnarray}
\end{subequations}
The localization length $L_{loc}$ can be derived via the Lyapunov exponent $\lambda$ \cite{TIM11,IKM11},
\begin{equation}\label{PRL-Lyap}
\frac{d}{L_{loc}}\equiv\lambda=-\frac{1}{2}\left\langle\ln\frac{d\theta_{n+1}}{d\theta_{n}}\right\rangle\,.
\end{equation}
Thus, the $\theta$-map \eqref{PRL-mapTheta-def} is the unique equation to be treated. We assume that the disorder is weak,
\begin{equation}\label{PRL-WeakDis}
\sigma^2\ll1\quad\mathrm{and}\quad(\sigma\varphi)^2\ll1,
\end{equation}
that allows us to develop a proper perturbation theory.

\emph{Phase distribution.} -- The phase distribution $\rho(\theta)$ can be found in the way described, e.g., in Refs.~\cite{IRT98,IKM11}. By expanding the exact $\theta$-map \eqref{PRL-mapTheta-def} up to the second order in perturbation and taking into account the uncorrelated nature of the disorder, see Eq.~\eqref{PRL-CorrDef}, one can obtain,
\begin{eqnarray}\label{PRL-theta}
\theta_{n+1}-\theta_n&=&-\gamma-\eta_a(n)U(\theta_n)\mp\eta_b(n)U(\theta_n-\gamma/2)\nonumber\\
&&-\sigma^2W(\theta_n),
\end{eqnarray}
where
\begin{subequations}\label{PRL-UW}
\begin{eqnarray}
&&U(\theta)=\varphi +\sin\varphi\cos(2\theta -\varphi)\,,\\
&&W(\theta)=\varphi[\cos(2\theta -2\varphi)\pm\cos(2\theta -2\gamma)]\nonumber\\
&&+\sin\varphi[\sin\theta\sin(\theta-\varphi)\pm\sin(\theta-\gamma/2)\sin(\theta-\varphi-\gamma/2)]\nonumber\\
&&+\sin^2\varphi\sin(4\theta-2\varphi-\gamma)\cos\gamma\,.
\end{eqnarray}
\end{subequations}
The unperturbed Bloch phase shift $\gamma$ over a unit $(a,b)$ cell is defined as
\begin{equation}\label{PRL-gamma}
\gamma=\varphi\pm\varphi=(1\pm 1)\omega d/2c,
\end{equation}
where ``plus" stands for the RH-RH structure, and ''minus" refers to the mixed RH-LH array. To proceed further, we pass from Eq.~\eqref{PRL-theta} to stationary Fokker-Plank equation for the phase distribution $\rho(\theta)$ ~\cite{IRT98,IKM11},
\begin{eqnarray}\label{PRL-eqFP}
&&\frac{d^2}{d\theta^2}\left[U^2(\theta)+U^2(\theta-\gamma/2)\right]\rho(\theta)\nonumber\\
&&+2\frac{d}{d\theta}\left[\frac{\gamma}{\sigma^2}+W(\theta)\right]\rho(\theta)=0,
\end{eqnarray}
which should be complemented by the normalization condition and the condition of periodicity $\rho(\theta+\pi)=\rho(\theta)$.

One can see that the form of solution of Eq.~\eqref{PRL-eqFP} strongly depends on whether the phase shift $\gamma$ is non-zero (RH-RH array) or vanishes (RH-LH array).

\emph{Homogeneous RH-RH array.} - In such a structure the Bloch phase \eqref{PRL-gamma} is finite, $\gamma=2\varphi=\omega d/c$, and for weak disorder the term in Eq.~\eqref{PRL-eqFP} containing $\gamma/\sigma^2$ prevails over the others for \emph{any} value of phase shift $\varphi$. Therefore, the phase distribution is \emph{uniform} within the first order of perturbation theory,
\begin{equation}\label{PRL-RhoUniform}
\rho(\theta)=1/\pi.
\end{equation}
The Lyapunov exponent can be derived by substitution of Eq.~\eqref{PRL-theta} into definition \eqref{PRL-Lyap}, and expanding the logarithm up to quadratic terms in disorder. After, the subsequent averaging over $\theta $ is trivial and one gets,
\begin{equation}\label{PRL-LyapHomogen}
d/L_{loc}\equiv\lambda=\sigma^2\sin^2\varphi\,;\qquad\qquad\varphi=\omega d/2 c .
\end{equation}
This gives standard $\omega-$dependence, $\lambda \propto \omega^2$ for  $\omega \rightarrow 0$,
\begin{equation}\label{PRL-LyapHomOmega}
\lambda=\sigma^2\omega^2 d^2/4c^2.
\end{equation}

\begin{figure}[t!!!]
\begin{center}
\includegraphics[scale=0.5]{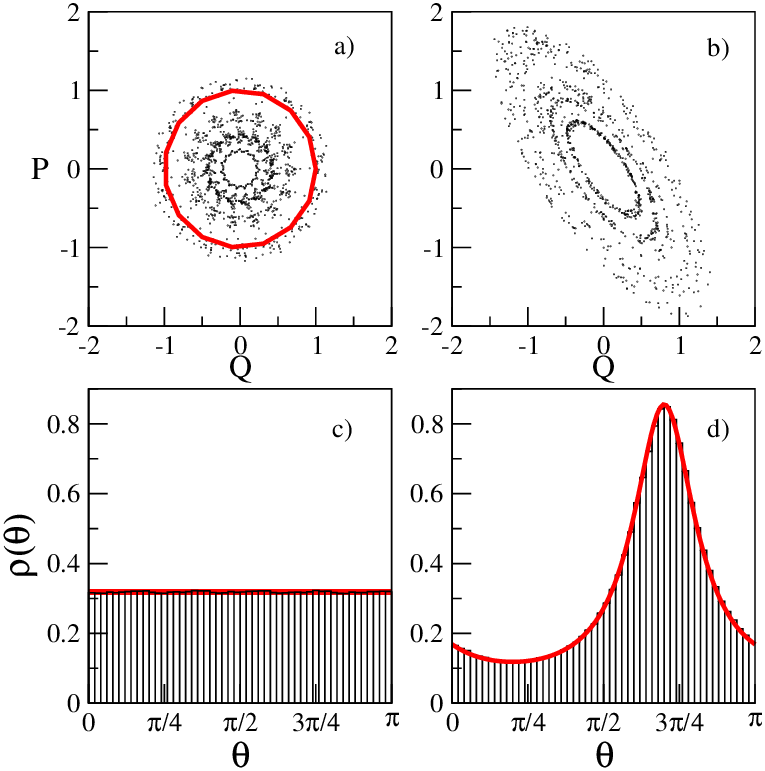}
\end{center}
\caption{(color online) a) phase space trajectory generated by Eq.~\eqref{PRL-mapQP} for RH-RH array with $N=10^4$, $\varphi=2\pi/30$, for zero disorder (solid circle), and for $\sigma^2=0.003$ (scattered points); b) one trajectory for RH-LH array with $N = 10^6$, $\varphi=2\pi/5$, $\sigma^2=0.003$. c) $\rho(\theta)$ from Eq.~\eqref{PRL-mapQP} for RH-RH array (histogram), and Eq.~\eqref{PRL-RhoUniform} (horizontal line); d) $\rho(\theta)$ from Eq.~\eqref{PRL-mapQP} for RH-RH array (histogram), and Eq.~\eqref{PRL-RhoNonUniform} (solid curve).}
\label{PRL-Fig01}
\end{figure}

\emph{Mixed RH-LH array.} -- The principally different situation emerges for the RH-LH array. In this case $\gamma=0$ \emph{independently} of the phase shift $\varphi$. As a result, we have $W(\theta)=-U(\theta)U'(\theta)$ in Eq.~(\ref{PRL-UW}), and Eq.~\eqref{PRL-eqFP} leads to a highly \emph{nonuniform} phase distribution,
\begin{equation}\label{PRL-RhoNonUniform}
\rho(\theta)=\frac{1}{\pi}\sqrt{\varphi^2 -\sin^2\varphi}\Big/U(\theta).
\end{equation}
Fig.~\ref{PRL-Fig01} displays perfect agreement between Eqs.~\eqref{PRL-RhoUniform}, \eqref{PRL-RhoNonUniform} and data obtained by the iteration of the exact map \eqref{PRL-mapQP}.

The above result means that to calculate the Lyapunov exponent via Eq.~\eqref{PRL-Lyap}, one needs to perform an average with the distribution $\rho(\theta)$ given by Eq.~(\ref{PRL-RhoNonUniform}). Surprisingly, the use of Eqs.~\eqref{PRL-theta} and \eqref{PRL-RhoNonUniform} results in the vanishing value of $\lambda$ \cite{TIM11}. Therefore, the Lyapunov exponent is determined by next orders of the perturbation theory.  However, the direct evaluation of high order terms in $\rho(\theta)$ is not possible with the above method \cite{T11}.

In order to proceed further, we suggest another method. The numerical data in Fig.~\ref{PRL-Fig01}b manifest that the trajectory has the form of fluctuating {\it ellipse} specified by angle with respect to axes, and by fixed aspect ratio. Therefore, one can expect that in new variables $\tilde{Q}_n, \tilde{P}_n$ obtained by rotating and rescaling the axes $Q,P$, the trajectory transforms into fluctuating {\it circle}. Thus, the distribution of a new phase $\Theta_n$ in the considered approximation will be uniform.

To follow this recipe, we rotate the $Q, P$- axes by angle $\tau$, with further rescaling the axes due to free parameter $\alpha$. In new coordinates the expressions \eqref{PRL-mapQP} and \eqref{PRL-QP-RTheta} -- \eqref{PRL-Lyap} conserve their forms, however, with the factors,
\begin{subequations}\label{PRL-ABCDnew}
\begin{align}
\tilde{A}_n&=A_n\cos^2\tau+(B_n-C_n)\sin\tau\cos\tau+D_n\sin^2\tau,\\
\tilde{B}_n\alpha^{-1}&=B_n\cos^2\tau-(A_n-D_n)\sin\tau\cos\tau+C_n\sin^2\tau,\\
\tilde{C}_n\alpha&=C_n\cos^2\tau+(A_n-D_n)\sin\tau\cos\tau+B_n\sin^2\tau,\\
\tilde{D}_n&=D_n\cos^2\tau-(B_n-C_n)\sin\tau\cos\tau+A_n\sin^2\tau,
\end{align}
\end{subequations}
instead of initial $A_n$, $B_n$, $C_n$, $D_n$.

\begin{figure}[t!!!]
\begin{center}
\includegraphics[scale=0.55]{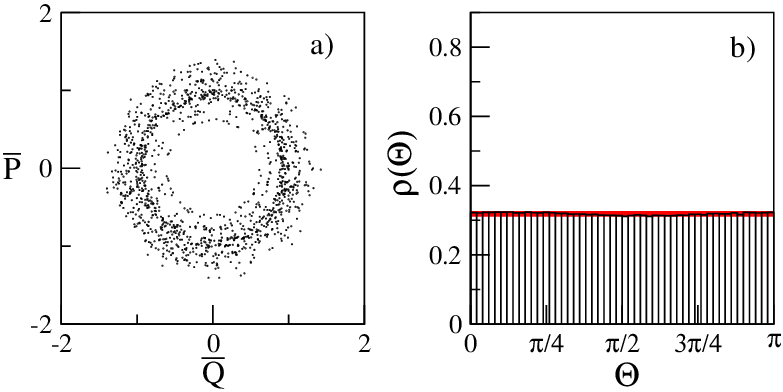}
\end{center}
\caption{(color online) (a) Phase space trajectory in new variables $(\tilde{Q},\tilde{P})$; (b) distribution $\rho(\Theta)$ generated by the transformed map with \eqref{PRL-ABCDnew} and \eqref{PRL-TauAlpha}, for $\gamma=0$, $\varphi=2\pi/5$, $\sigma^2=0.02$ and $N=10^7$.}
\label{PRL-Fig02}
\end{figure}

Now the distribution $\rho(\Theta)$ for new phase $\Theta$ can be found starting from the quadratic expansion of Eq.~\eqref{PRL-mapTheta-def} with new coefficients \eqref{PRL-ABCDnew} and $\gamma=0$,
\begin{equation}\label{PRL-ThetaNew}
\Theta_{n+1}-\Theta_{n}=[\eta_a(n)-\eta_b(n)]V(\Theta_n)+\sigma^2 V(\Theta_n)V'(\Theta_n).
\end{equation}
Here the function $V(\Theta)$ is defined by
\begin{eqnarray}\label{PRL-V}
&&V(\Theta)=\sin\varphi\sin(2\tau-\varphi)\sin2\Theta\nonumber\\
&&+\frac{\alpha}{2}[\varphi -\sin\varphi\cos(2\tau-\varphi)][\cos2\Theta-1]\nonumber\\
&&-\frac{1}{2\alpha}[\varphi+\sin\varphi\cos(2\tau-\varphi)][\cos2\Theta+1].
\end{eqnarray}
The stationary Fokker-Plank equation corresponding to $\Theta$-map \eqref{PRL-ThetaNew} reads,
\begin{equation}\label{PRL-eqFPnew}
\frac{d}{d\Theta}\left[V^2(\Theta)\frac{d}{d\Theta}\rho(\Theta)+V(\Theta)V'(\Theta)\rho(\Theta)\right]=0.
\end{equation}
From this equation one gets that the phase distribution is uniform, $\rho(\Theta)=1/\pi$, and the trajectory is, indeed, a fluctuating circle provided that
\begin{equation}\label{PRL-UniformCond}
\frac{d}{d\Theta}{V(\Theta)V'(\Theta)}=0.
\end{equation}
With the use of Eqs.~\eqref{PRL-V} and \eqref{PRL-UniformCond} we now can obtain the desired expressions for the angle $\tau$, parameter $\alpha$ and function $V(\Theta)$ (which is actually no more $\Theta$-dependent),
\begin{equation}\label{PRL-TauAlpha}
\tau=\frac{\varphi}{2},\quad\alpha^2 =\frac{\varphi+\sin\varphi}{\varphi-\sin\varphi},\quad V(\Theta)=\sqrt{\varphi^2-\sin^2\varphi}.
\end{equation}
The data presented in Fig.~\ref{PRL-Fig02} confirm the success of our approach: in new variables the trajectory is a fluctuating circle and the phase distribution is uniform.

The Lyapunov exponent $\lambda$ can be now obtained via Eq.~(\ref{PRL-Lyap}) with the change $\theta_n \rightarrow \Theta_n$. Taking into account that $\lambda$ vanishes within quadratic approximation in disorder, we expand the $\Theta$-map of the form \eqref{PRL-mapTheta-def} with the coefficients \eqref{PRL-ABCDnew} up to the fourth order in perturbation. By substituting the resulting expression into Eq.~\eqref{PRL-Lyap} and expanding the logarithm within the same approximation, after the averaging over $\Theta_n$ with flat distribution, we arrive at final expression,
\begin{equation}\label{PRL-LyapMixed}
\frac{d}{L_{loc}}=\frac{\zeta+2}{4}\sigma^4\frac{[(2\varphi^2-\sin^2\varphi)\cos\varphi-\varphi\sin\varphi]^2}{\varphi^2-\sin^2\varphi}.
\end{equation}
Here $-2\leqslant\zeta<\infty$ stands for the excess kurtosis, $\zeta=\langle\eta(n)^4\rangle/ \langle\eta(n)^2\rangle^2-3 $,  the constant specified by the form of distribution of $\eta_{a,b}(n)$. For Gaussian and flat distributions we have $\zeta=0,-6/5$, respectively.

Eq.~\eqref{PRL-LyapMixed} determines the asymptotics for small $\varphi$,
\begin{equation}\label{PRL-LyapMixedOmega}
\frac{d}{L_{loc}}\equiv\lambda=\frac{2^4}{3^35^2}(\zeta+2)\sigma^4\varphi^8 \quad\mathrm{for}\ \varphi\ll1\ll\sigma^{-1},
\end{equation}
that results in a quite surprising frequency dependence of the Lyapunov exponent, $\lambda\propto\omega^8$. Thus, the dependence $\lambda\propto\omega^6$, numerically found for small $\omega$ in Refs.~\cite{Ao07,Ao10} should be treated as the intermediate one, apparently emerging due to not sufficiently large lengths $N$ over which the average of $\lambda$ is performed.

\begin{figure}[t!!!]
\begin{center}
\includegraphics[scale=0.4]{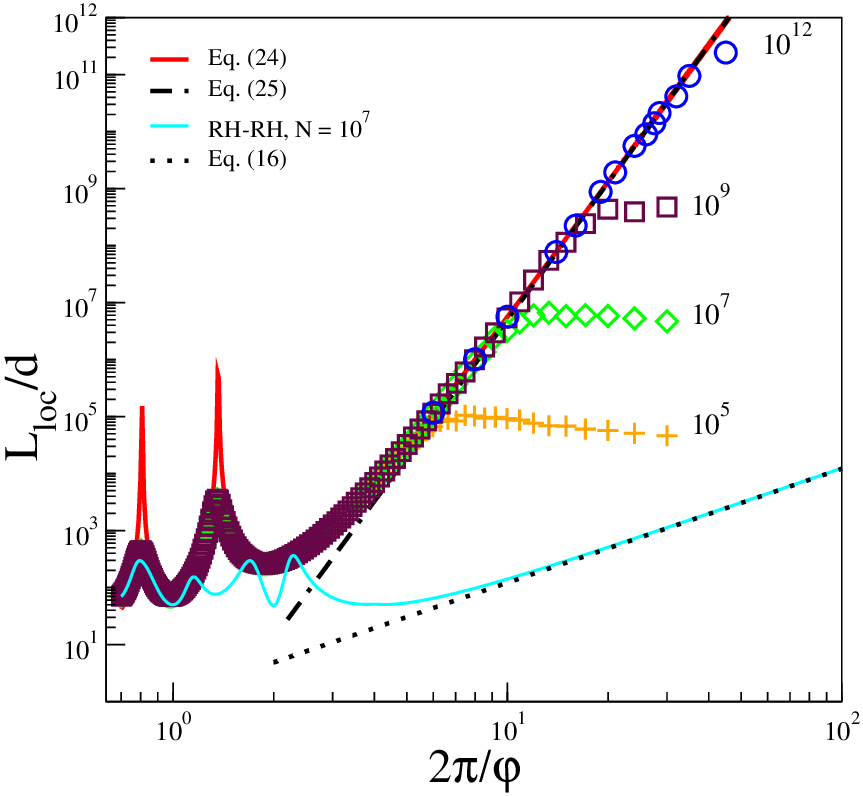}
\end{center}
\caption{(color online) Localization length for RH-LH and RH-RH arrays of different lengths $N$ versus the normalized wavelength $2\pi/\varphi=4\pi c/\omega d$, for $\sigma^2=0.02$.}
\label{PRL-Fig03}
\end{figure}

Figure \ref{PRL-Fig03} shows excellent agreement between numerical data obtained for the localization length from exact Eq.~\eqref{PRL-Lyap} and Eqs.~(\ref{PRL-LyapMixed}), (\ref{PRL-LyapMixedOmega}) for RH-LH, as well as Eqs.~(\ref{PRL-LyapHomogen}), (\ref{PRL-LyapHomOmega}) for RH-RH arrays with $\zeta =-6/5$. For sequence lengths $N=10^5; 10^7$ and $10^9$ the data are obtained with ensemble averaging over 100 realizations of disorder, while at $N=10^{12}$ only one realization is used.

\emph{Conclusions.} -- Our approach allows one to resolve the problem of non-conventional Anderson localization for bilayered RH-LH arrays with equal widths, $d_a=d_b$, and derive Eq.~(\ref{PRL-LyapMixed}) for the localization length. As we show, the peculiarity of this model is entirely due to zero value of unperturbed Bloch phase, $\gamma=0$. We derive the expression for the Lyapunov exponent which appears to be defined by the fourth order of perturbation theory in disorder, and valid for {\it any} value of frequency $\omega$. Our results prove that for small $\omega$  the localization length is enormously  large, $L_{loc} \propto \sigma^{-4} \omega^{-8}$.

According to the analysis given above, one can conclude that the non-conventional dependence $L_{loc} \propto \sigma^{-4} \omega^{-8}$ also emerges when $d_a \neq d_b$, in case of equal unperturbed optic lengths, $n_a d_b = |n_b d_b|$, provided that the impedances are also equal, $Z_a=Z_b$. Another generalization is due to Eq.~(\ref{PRL-theta}), according to which the same dependence for $L_{loc} $ is expecting to occur when the difference $\Delta = n_a d_b - |n_b d_b|$ is sufficiently small, $\Delta \ll \sigma ^2$.

Our results contribute to the theory of bilayered structures with an inclusion of left-handed materials, and may be used in experimental realizations of structures with specific properties of transmission.

\emph{Acknowledgments.}-- F.M.I acknowledges support from VIEP grant EXC08-G of the BUAP (Mexico).



\end{document}